\documentstyle[epsf]{article}
\def\t{\hbox}

\def\f{\frac}

\def\o{\over}
\def\q{\quad}
\def\p{\varphi}

\def\D{\partial}

\def\k{\kappa}

\def\a{\alpha}

\def\d{\delta}
\def\be{\begin{equation}}
\def\ee{\end{equation}}
\def\bea{\begin{eqnarray}}
\def\eea{\end{eqnarray}}
\def\ba{\begin{array}}
\def\ea{\end{array}}

\def\pr{\prime}
\def\pa{\partial}
\begin{document}

\hspace{1.5cm} Physics Letters A254, N3-4, (1999), pp.141-148. 
\vspace{0.5cm}

\centerline{\bf EXAMPLE OF TWO DIFFERENT POTENTIALS WHICH HAVE}
\centerline{\bf PRACTICALLY THE SAME FIXED-ENERGY PHASE SHIFTS }

\vspace{0.5cm}

\centerline{\footnotesize
Ruben G. AIRAPETYAN
\footnote{E-mail: airapet@math.ksu.edu},\quad
Alexander G. RAMM
\footnote{E-mail: ramm@math.ksu.edu, fax 785-532-0546, 
please 
send correspondence to this
 author}, \quad
and \quad Alexandra B. SMIRNOVA
\footnote{E-mail: smirn@math.ksu.edu}
}
\vspace{0.3cm}
\centerline{{\footnotesize\it Department of Mathematics, Kansas State University,}}
\centerline{{\footnotesize\it Manhattan, Kansas 66506-2602, U.S.A.}}

\vspace{1cm}

{\footnotesize
The Newton-Sabatier procedure for finding the potential
from fixed-energy phase shifts is analyzed.
A method is proposed for finding two quite different
 spherically-symmetric
real-valued, piecewise-constant, compactly supported potentials 
which generate at a fixed energy
the phase shifts $\delta_{\ell}$
 which are practically indistinguishable for all $\ell$.
In particular, an explicit concrete example of two
such potentials $q_j,$ $j=1,2,$ is demonstrated. These potentials
have the properties:

$1)\q \sup |q_1-q_2|>3,\t{ and }\q q_j,\q j=1,2, 
\t{ are of order of magnitude 1,}$

$2)\q \delta_l^{(1)}=\delta_l^{(2)}\t{ for }l=0,\dots,4\t{ and } 
|\delta_l^{(1)}-\delta_l^{(2)}|\leq 10^{-5},\q l>4.$

PACS: 0380, 0365.

}



\section{Introduction}          
\vspace*{-0.5pt}
\noindent
\setcounter{equation}{0} 
\renewcommand{\theequation}{1.\arabic{equation}}

Finding real-valued potentials $q(r)$, $r=|x|$, from the set
$\{\delta_l\}_{l=0,1,...}$ of the fixed-energy phase shifts 
(say at $k=1$), is an 
inverse problem which was much discussed in the literature 
\cite{ael}, \cite{cs}, \cite{l}, \cite{n}, \cite{r1}, \cite{r2}.
Physicists worked much on this problem. They used a parameter
fitting procedure
\cite{n}, \cite{cs} which consists of the following steps:
\begin{description}
\item[Step 1.] Given  phase shifts $\delta_l$ at a fixed
energy, which we take to be $k^2=1$ without loss of generality, one
solves an infinite system of linear algebraic equations for some
coefficients $c_l$ (system (12.2.7) in \cite{cs}; the references
to formulas with such numeration here and below are
given alawys to the formulas from [2]).
\item[Step 2.] Given these coefficients $c_l,$ one constructs
some symmetric kernel ((12.2.1) in  and solves linear
integral     
equation (12.1.2) for the kernel $K(r,\rho).$ If
$K(r,\rho)$
is found, then the potential $q(r)$ is calculated by the formula
(12.2.11):
\end{description}   
\be\label{eq11}
q(r)=-\f{2}{r}\, \f{d}{dr}\, \f{K(r,r)}{r}.
\ee

This procedure is presented in [5] and [2] as an inversion method and
many physicists consider it as one.
In this introduction we analyze this procedure.

Let us define an inversion method as a method
which satisfies the following conditions: 

1) given the exact data,
corresponding to a $q(r)\in Q$, where $Q$ is a class
of potentials in which the inverse scattering problem is
uniquely solvable, one can carry through theoretically
(and not necessarily computationally)  each step of
the method  (we neglect here the numerical errors
unavoidable in the numerical implementation of any method), 

2) the method yields the unique solution $q\in Q$ to the inverse problem.

In this paper we take the class $Q$ of
compactly supported real-valued
 potentials $q(r)$ which satisfy the conditions:
 $\int_0^a |q(r)|^2 r^2 dr<\infty,\,\, q(r)=0$ for $r>a$,
 where $a>0$ is an arbitrary large fixed number
 which is called the range of the potential.
  For $q\in Q$ the inverse
 scattering problem with fixed-energy phase shifts as the scattering
 data has unique solution [7]. Let us assume that the
 given scattering data
 come from a potential in this class. The question is:
 can one reconstruct such a potential using the
 Newton-Sabatier (NS) procedure
 and assuming that this procedure can be carried through?
 We prove that the answer to the above question is no.

Solving the inverse scattering problem
includes the requirement that the potential obtained by an inversion
method produces the original data from which
it was reconstructed. In [2] this property
is called consistency of the method.

The basic results of our paper are:

1) it is shown that if one applies the NS procedure
to the data generated by a potential $q\in Q$, then,
assuming that this procedure can be carried through, it
yields the potentials $q(r)$  which are analytic in a neighborhood
of the positive semi-axis of the complex $r-$plane and may have a simple
pole at the origin, and therefore is not equal to the potential
$q\in Q$ with which we started.

2) it is shown that the stability estimate ( see estimate (1.5) below)
from [8] is
practically accurate: an example is constructed
of two different piecewise-constant compactly
supported potentials which produce practically the same (up to the error
of order $10^{-5}$ ) fixed-energy phase shifts for all $\ell$.

Therefore even if
 both steps in the NS method with exact data,
 corresponding to a $q\in Q$, can be
carried through in the above class $Q$,
they lead to a potential $q_1(r)\notin Q$,
 which has very special property (P),
namely the function $rq(r)$ has to be analytic in a neighborhood
of the positive semiaxis of the complex $r-$plane.
This property
the potentials from class $Q$ do not have.

Consistency of the NS procedure
 is not proved in [2] and [5]:  assuming that the NS procedure
can be carried through for the given data,
there is no guarantee that the potential generated
by this procedure produces the same data with which one started.

In [2, p.205] an argument is given to prove the consistency of the NS
procedure, but this argument is not
completely convincing because: 1) it is not quite clear why
$A'_{\ell}$ and $\delta_{\ell}'$ solve equation (12.2.5),
2) existence and uniqueness of the solution to Goursat-type
problem (11.2.5)-(11.2.7) are not proved in [2];
since the coefficients in this problem are variable and
have non-integrable singularity at the origin, the
desired uniqueness and existence do not follow from the known results;
in the literature one can find results on the existence and uniqueness of
the transformation kernels for equation (11.1.2), but these
kernels depend on $\ell$, while the transformation kernel
used in [2] and [5] should be independent on $\ell$.

Note that the {\it basic ansatz of the NS procedure} says that the 
transformation kernel (12.2.3) is generated by the symmetric
kernel (12.2.1) and condition (12.2.2) holds. 

{\it There is no guarantee that this ansatz can yield 
 (by solving system (12.2.4) and then using formula (12.2.11)) a
potential which can produce generic scattering data.
Indeed, as we prove below, the potential
which is obtained by the NS procedure (in the case when this procedure
can be carried through) has to satisfy property (P),
which generic potentials do not have. Therefore the usage of this ansatz is
not fully justified: there is no guarantee that one can fit the exact
phase shifts, generated by a potential $q\in Q$, using the potentials
produced by the NS procedure based on the above ansatz.}

In the discussion of these principal points we are not talking at all
about the difficulties which come from the fact that in practice the
 data are never exact.

There is no guarantee that the 
steps 1 and 2 of the NS procedure with exact data can be carried through,
 in general, because:

1) System (12.2.7) may have, in general, no solution in the class (12.2.2),

2) Integral equation (12.1.2) for the kernel
 $K(r,\rho)$ may be not solvable for some 
(not too small) $r>0$.

Note that existence of the transformation kernel (11.2.4)
is not proved in [2] and [5] but was proved recently in [11].

  It is easy to prove that equation (12.1.2)
  is uniquely solvable for sufficiently small $r>0$ since in this
  case the norm of the corresponding integral operator in $C(0,r)$
  tends to $0$ as $r\to 0$.

 If it is possible to solve (12.2.7) in the class (12.2.2) and
 equation  (12.1.2) is uniquely solvable for all $r>0$,
then the potential (12.2.11) must have the following special property:

(P) {\sl The function $rq(r)$ must be an analytic function of $r$
in a neighborhood of the  positive semiaxis $(0, \infty)$}.

The proof of property (P) for the potentials obtained by the NS scheme
is given below.

From the logical point of view property (P) is sufficient for
the proof of our basic conclusion:

{\sl Conclusion: For the generic data corresponding
to a compactly supported potential $q(r)\in Q$  the NS
procedure cannot reproduce the original potential.}

Thus, we have the following logical possibilities:

{\it i) Either the NS procedure can be carried through
for the phase shifts corresponding to a $q\in Q$
and the reconstructed by this procedure potential $q_1(r)$
reproduces the original phase shifts,
and then the inverse scattering problem has more than one
solution and the solution $q\in Q$ cannot be obtained
by the NS procedure;

ii) or the NS procedure cannot be carried through
for the phase shifts corresponding to $q\in Q$;

iii) or the NS procedure can be carried through for these phase shifts but
the obtained potential $q_1(r)$ does not generate the original
phase shifts, in which case the NS procedure does not solve
the inverse scattering problem.}

{\it Similar argument holds for 
 any class of potentials which do not have property (P) and
for which the uniqueness of the solution to the inverse scattering
problem with exact data is valid.}

Let us now prove that the potential, obtained by the NS procedure,
assuming that this procedure can be carried through
for the given data, must have property (P). 

{\bf Proof of property (P).}

The functions $u_{\ell}$ in (12.2.3) are entire functions of $r$ which
have the following known asymptotics for large $\ell$: 
$u_{\ell}\sim \sqrt {\frac r2} (\frac {er}{2\ell +1})^{\frac {2\ell +1}{2}}
 (2\ell +1)^{-1/2}$.
Therefore the function (12.2.1),
with $c_{\ell}$ satisfying (12.2.2), is an entire
function of $r$ and $r'$.

Let us first prove that $rq(r)$ is analytic in a neighborhood of the
origin, since this claim already is sufficient for a justification of our basic
conclusion and
the proof of this claim is quite easy.

The system (12.2.4) has the matrix $L_{\ell \ell'}(r)=\int_0^r
u_\ell(\rho) u_{\ell'}(\rho) \rho^{-2}d\rho$ (see formula (11.4.5)
with $V_0=0$). Since $u_\ell(r)$ is an entire function of $r$ and
is $O(r^{\ell +1})$ as $r\to 0$, one concludes that $L_{\ell \ell'}(r)$
is an entire function of $r$. For sufficiently small $r$
system (12.2.4) has dominating diagonal terms and therefore
is uniquely solvable by iterations. Since $u_\ell(r)$ and
$L_{\ell \ell'}(r)$ are
entire functions of $r$, the Neumann series
yields the solution $\varphi_\ell(r)$
 of (12.2.4) which is an analytic function
of $r$ in a neighborhood of the point $r=0$. This implies that the
kernel $K(r,r'),$ defined by (12.2.3), is analytic with
respect to both variables in a neighborhood of $r=0,\, r'=0.$
Therefore $K(r,r)$ is analytic in a neighborhood of $r=0$ and
consequently the potential (12.2.11) (defined by formula (1.1))
is analytic in a punctured at $r=0$ neighborhood of the origin
and the function $rq(r)$ is analytic in a neighborhood of the origin.
Our first claim is proved.

Let us now prove that $rq(r)$ is analytic in a neighborhood
of the positive semiaxis provided that equation (12.1.2)
is uniquely solvable. The kernel and the free term
 of this equation, as we have proved above,
are entire functions of $r$ and $r'$. It is known that the solution
of a Fredholm second kind integral equation whose kernel and free term
depend on a parameter $z$ analytically in a certain region $\Delta$
of a complex plane, is analytic in a neighborhood of a
point $z_0\in \Delta$, provided that
the above Fredholm equation is uniquely solvable at $z=z_0$ (see [10]
for a more general result). 

Let us
introduce the new variables $\rho=rs, r'=rt$
in equation (12.1.2) with the aim to get equation with a fixed
integration region. Equation (12.1.2) becomes:
$$K(r,rt)=f(r,rt)-\int_0^1 K(r,rs)f(rs, rt)r^{-1}s^{-2}ds.$$
Let 
$$a(s,t,r):=f(rs,rt)(rst)^{-1},\,\,  K(r,rs)s^{-1}:=b(s,r),\,\,
f(r,rt)t^{-1}:=g(r,t).$$
Then the above equation takes the form: 
$$ b(t,r)=g(r,t)-\int_0^1 a(s,t,r)b(s,r)ds.
$$ 
This equation is equivalent to equation
(12.1.2) in [2]. It is a Fredholm second
kind equation whose kernel $a(s,t,r)$ and the free term $g(r,t)$
depend on the parameter $r$ analytically on the whole
complex $r-$plane because the kernel $f(s,t)$ is an entire function
of $s$ and $t$, as we proved above, and it vanishes at $s=0$ and at $t=0$.
Also $g(r,t)$ and  $a(s,t,r)$  are entire functions of $t$.
If the NS procedure can be carried through,
then this equation has to be uniquely
solvable for any $r>0$. Therefore its solution is an analytic
function of $r$ in a neighborhood of positive semiaxis $(0, \infty)$
and an analytic function of $t$ on the whole complex $t-$plane.
This implies that the kernel $K(r,r)$ is an analytic 
function of $r$ in a neighborhood
of the positive semiaxis $(0, \infty)$.
 Thus, the potential (12.2.11) has property (P).
The proof is complete.
$\Box$
 
Therefore, in general,  
the inverse scattering problem is not solvable
 in the class $Q$ by means of the NS procedure.

The
uniqueness of the recovery of $q(r)$ from the fixed-energy phase shifts
$\delta_l$, $l\ge 0$, is not established in \cite{cs}. Such a uniqueness
theorem follows for $q\in Q$
from the uniqueness
theorem proved in \cite{r2} for potentials which are not
necessarily  spherically symmetric.

No such theorem is known for potentials 
which decay as some power of $\f{1}{r},$
that is,     
\be\label{eq12}
|q|=O\left(\f{1}{(1+r)^m}\right),\q m>2 \t{ is fixed}.
\ee
It is known (see \cite{r4}) that the range $a$ of a compactly supported
potential, which does not change sign in some neighborhood
$(a-\varepsilon , a)$, can be
calculated in terms of $\{\delta_l\}:$
\be\label{eq13}
a=\lim_{l\to \infty}\left(\f{2l+1}{e}|\delta_l|^{\f{1}{2l}}\right).
\ee
It follows from  (\ref{eq13}) that
\be\label{eq14}
|\delta_l|\leq c\left(\f{ae[1+o(1)]}{2l+1}\right)^{2l},\q c=const >0.
\ee
This implies that $\delta_l$ decay very fast as $l\to \infty$, provided that  
$q(r)$ is compactly supported.

If $q(r)$ decays on a power scale  (\ref{eq12}) then $\delta_l$ decays also
as some positive power of $l^{-1}$ (see \cite{ll}).

In this paper we construct two piecewise-constant, real-valued, 
compactly supported potentials $q_1(r)$ and $q_2(r),$ which are quite different
(see Table 1 in sec.3) and which generate practically identical phase shifts
$\delta_l.$ 

The potentials $q_j,$ $j=1,2,$ are constructed by a computer code, 
which recovers $q(r)$ (in the above class of potentials) from
the data $\{\delta_l\},\,l\geq 0.$

It is known that the inverse problem of quantum scattering with fixed-energy
scattering data is very ill-posed, if the energy is not very high,
and the  numerical reconstruction of the potential
from noisy scattering data is quite difficult.

However, our numerical results also show that if an initial approximation to
the unknown potential is known with relatively small error ($\approx 10^{-1}$),
then the recovery of $q(r)$ is possible with high accuracy ($ 10^{-14}$).

In \cite{r3} a mathematically
justified method is proposed for stable inversion
of fixed-energy scattering data
for not necessarily spherically symmetric potentials
with compact support. 
If the potential $q(r)$ is bounded spherically , symmetric and
has compact support then the result in \cite{r3} yields an algorithm for
reconstruction of  
$q(r)$ from the phase shifts $\delta_l$  known with some error
$\varepsilon$. This algorithm 
yields $\hat q_{\varepsilon}$ such that
\be\label{eq15}
\sup_{\xi}|\tilde q(\xi)
-\hat q_{\varepsilon}|\leq c\f{(\ln|\ln\varepsilon |)^2}{|\ln\varepsilon
|},\q \varepsilon \to 0, 
\ee
where $c>0$ is some constant which does not depend on $\varepsilon$, but does depend
on some norm of the potential, $\tilde q(\xi)$ is the Fourier transform
of $q(x)$.
The right-hand side of (\ref{eq15}) tends to zero
as $\varepsilon \to 0$, however it tends very slowly. Therefore practically one can
hope to recover $q(r)$ from the fixed-energy scattering data, that is, from the phase
shifts $\delta_l$ at a fixed $k,$ only if either the class of the potentials is a
priori chosen to be rather narrow, for instance, finite-parametric, or the accuracy
of the data is very high. Our numerical results confirm these conclusions. 
The example we constructed is an illustration of the stability estimate (1.5).

In section 2 a numerical method is described which we use to invert the
fixed-energy phase shifts for the potential $q(r).$ In section 3 the numerical results
are presented.

\section{Numerical method}          
\vspace*{-0.5pt}
\noindent
\setcounter{equation}{0}
\renewcommand{\theequation}{2.\arabic{equation}}

Consider a finite set of points $0=r_0< r_1 <r_2<\dots<r_N=R$
and a piecewise-constant potential
\be\label{pot}
q(r)=q_i,\t{ on } [r_{i-1},r_i) \t{ for } i=1,\dots, N, \t{ and } q=0
\t{ for }r\ge R. 
\ee
Denote $\k_i^2:=k^2-q_i$, where $i=1,\dots, N,$
and $k$ is some fixed positive number.
Consider the following problem for the radial Schr\"odinger equation:
\be
\f{d^2\p_l}{dr^2}+\Biggl(k^2-\f{l(l+1)}{r^2}\Biggl)\p_l=q\p_l,
\q \lim_{r\to 0}(2l+1)!!r^{-l-1}\p_l(r)=1,
\ee
which we rewrite as:
\be\label{sc}
\f{d^2\p_l}{dr^2}+\Biggl(\k_i^2-\f{l(l+1)}{r^2}\Biggl)\p_l=0
\ee
on the interval $r_{i-1}\le r < r_i$.
On $[r_{i-1},r_i)$ one has the following general solution of (\ref{sc})
\be
\p_l(r)=A_iu_l(\k_ir)+B_iv_l(\k_ir),
\ee
where
$$
u_l(z)=\sqrt{\pi z\o 2}J_{l+1/2}(z),\q v_l(z)=\sqrt{\pi z\o 2}Y_{l+1/2}(z).
$$
We assume below that $\k_i$ does not vanish for all $i$. If $\k_i=0$
for some $i$ then our approach is still valid with obvious changes.

From the regularity of $\p_l$ at zero one gets $B_1=0$. Denote
$x_i=B_i/A_i$, then $x_1=0$. 
We are looking for the continuously differentiable solution $\p_l$.
Thus, the following interface conditions hold:
\be\ba{lcc}
A_iu_l(\k_ir_i)+B_iv_l(\k_ir_i)=A_{i+1}u_l(\k_{i+1}r_{i})+
B_{i+1}v_l(\k_{i+1}r_{i}),\\ \\
\f{\k_i}{\k_{i+1}}[A_iu_l^\pr(\k_ir_i)+B_iv_l^\pr(\k_ir_i)]=
A_{i+1}u_l^\pr(\k_{i+1}r_{i})+B_{i+1}v^\pr_l(\k_{i+1}r_{i}).
\end{array}\ee
The Wronskian $W(u_l(r),v_l(r))=1$, thus
\be\ba{lcc}
A_{i+1}=v_l^\pr(\k_{i+1}r_{i})[A_iu_l(\k_ir_i)+B_iv_l(\k_ir_i)]
-\f{\k_i}{\k_{i+1}}v_l(\k_{i+1}r_{i})[A_iu_l^\pr(\k_ir_i)+B_iv_l^\pr(\k_ir_i)],
\\
\\
B_{i+1}=\f{\k_i}{\k_{i+1}}u_l(\k_{i+1}r_{i})[A_iu_l^\pr(\k_ir_i)+
B_iv_l^\pr(\k_ir_i)]
-u_l^\pr(\k_{i+1}r_{i})[A_iu_l(\k_ir_i)+B_iv_l(\k_ir_i)].
\ea\ee
Therefore
\be
\pmatrix{A_{i+1}\cr B_{i+1}}=\f{1}{\k_{i+1}}\pmatrix{\a^i_{11} & \a^i_{12}\cr
\a^i_{21} & \a^i_{22}}\pmatrix{A_{i}\cr B_{i}},
\ee
where the entries of the matrix $\alpha^i$ can be written explicitly:
\be\ba{lcc}
\a^i_{11}=\k_{i+1}u_l(\k_ir_i)v_l^\pr(\k_{i+1}r_{i})-
\k_{i}u_l^\pr(\k_ir_i)v_l(\k_{i+1}r_{i}),
\\
\\
\a^i_{12}=\k_{i+1}v_l(\k_ir_i)v_l^\pr(\k_{i+1}r_{i})-
\k_{i}v_l^\pr(\k_ir_i)v_l(\k_{i+1}r_{i}),
\\
\\
\a^i_{21}=\k_{i}u_l^\pr(\k_ir_i)u_l(\k_{i+1}r_{i})-
\k_{i+1}u_l(\k_ir_i)u_l^\pr(\k_{i+1}r_{i}),
\\
\\
\a^i_{22}=\k_{i}v_l^\pr(\k_{i}r_i)u_l(\k_{i+1}r_{i})-
\k_{i+1}v_l(\k_ir_i)u_l^\pr(\k_{i+1}r_{i}).
\ea\ee
Thus
\be\label{xk}
x_{i+1}=\f{\a^i_{21}+\a^i_{22}x_i}{\a^i_{11}+\a^i_{12}x_i},\q
x_i:=\f{B_i}{A_i}
\ee
Denote the phase shift $\d(l,k)$ by the formula
\be
\p_l(r)\sim{|F(l,k)|\over k^{l+1}}
\sin(kr-\frac{\pi l}{2}+
\delta(l,k))\quad r\to\infty\enspace ,
\ee
where $F(l,k)$ is the Jost function.
For $r>R$
\be\label{as}
\p_l(r)=A_{N+1}u_l(kr)+B_{N+1}v_l(kr).
\ee
From (\ref{as}) and the asymptotics
$u_l(kr)\sim\sin(kr-l\pi/2),\q v_l(kr)\sim-\cos(kr-l\pi/2)$,
$r\to\infty$, one gets:
\be\label{dkl}
\tan\delta(k,l)=-\f{B_{N+1}}{A_{N+1}}=-x_{N+1},\q
\delta(k,l)=-\arctan x_{N+1}.
\ee
Now let us assume that the column vector 
${\bf\d^*}=(\d_0^*,\dots,\d_{N-1}^*)$ is given
and  introduce the nonlinear operator $\Phi : R^N\to R^N$,
which maps a column vector ${\bf\k}=(\k_1,\dots,\k_N)$ into a vector
$\Phi({\bf\k})={\bf\d}-{\bf\d^*}$, where ${\bf\d}$ is given by (\ref{dkl}) 
and $\d_j^*,\q j=0,\dots,N-1$, are given fixed energy phase shifts.
Thus, we are looking for the solution to the problem
\be\label{equat}
\Phi({\bf\k})=0.
\ee
In order to solve this equation by means of the Newton method, first one
has to find the Jacobian 
$||\Phi^\pr||=||\D_j\Phi_i||$, where  $\pa_j=\pa/\pa\k_j$.
 
Since $\a^i=\a^i(\k_i,\k_{i+1})$, $x_i=x_i(\k_1,\dots,\k_i)$, one has
\be
\pa_jx_{i+1}=\f{\t{det}\a^i}{(\a^i_{11}+\a^i_{12}x_i)^2}\pa_jx_i,\q j<i,
\ee
and, for $j=i$ or $j=i+1$, one gets:
$$
\pa_jx_{i+1}=\f{\t{det}\a^i}{(\a^i_{11}+\a^i_{12}x_i)^2}\pa_jx_i+
$$
\be
+\f{(\a^i_{11}+\a^i_{12}x_i)(\pa_j\a^i_{21}+x_i\pa_j\a^i_{22})
-(\a^i_{21}+\a^i_{22}x_i)(\pa_j\a^i_{11}+x_i\pa_j\a^i_{12})}
{(\a^i_{11}+\a^i_{12}x_i)^2},
\ee
where
$$\ba{lll}
\pa_i\a^i_{11}=\k_{i+1}r_iu_l^\pr(\k_ir_i)v_l^\pr(\k_{i+1}r_{i})-
u_l^\pr(\k_ir_i)v_l(\k_{i+1}r_{i})
-\k_{i}r_iu_l^{\pr\pr}(\k_ir_i)v_l(\k_{i+1}r_{i}),
\\
\\
\pa_i\a^i_{12}=\k_{i+1}r_iv_l^\pr(\k_ir_i)v_l^\pr(\k_{i+1}r_{i})-
v_l^\pr(\k_ir_i)v_l(\k_{i+1}r_{i})
-\k_{i}r_iv_l^{\pr\pr}(\k_ir_i)v_l(\k_{i+1}r_{i}),
\\ 
\\
\pa_i\a^i_{21}=u_l^{\pr}(\k_ir_i)u_l(\k_{i+1}r_{i})
+\k_ir_iu_l^{\pr\pr}(\k_ir_i)u_l(\k_{i+1}r_{i})
-\k_{i+1}r_iu_l^\pr(\k_ir_i)u_l^\pr(\k_{i+1}r_{i}),
\\ 
\\
\pa_i\a^i_{22}=v_l^{\pr}(\k_ir_i)u_l(\k_{i+1}r_{i})
+\k_ir_iv_l^{\pr\pr}(\k_ir_i)u_l(\k_{i+1}r_{i})
-\k_{i+1}r_iv_l^\pr(\k_ir_i)u_l^\pr(\k_{i+1}r_{i}),
\\ 
\\
\pa_{i+1}\a^i_{11}=u_l(\k_ir_i)v_l^\pr(\k_{i+1}r_{i})+
\k_{i+1}r_iu_l(\k_ir_i)v_l^{\pr\pr}(\k_{i+1}r_{i})-
\k_{i}r_iu_l^{\pr}(\k_ir_i)v_l^\pr(\k_{i+1}r_{i}),
\\
\\
\pa_{i+1}\a^i_{12}=v_l(\k_ir_i)v_l^\pr(\k_{i+1}r_{i})+
\k_{i+1}r_iv_l(\k_ir_i)v_l^{\pr\pr}(\k_{i+1}r_{i})-
\k_{i}r_iv_l^{\pr}(\k_ir_i)v_l^\pr(\k_{i+1}r_{i}),
\\
\\
\pa_{i+1}\a^i_{21}=\k_{i}r_iu_l^{\pr}(\k_ir_i)u_l^{\pr}(\k_{i+1}r_{i})
-u_l(\k_ir_i)u_l^\pr(\k_{i+1}r_{i})-
\k_{i+1}r_iu_l(\k_ir_i)u_l^{\pr\pr}(\k_{i+1}r_{i}),
\\
\\
\pa_{i+1}\a^i_{22}=\k_{i}r_iv_l^{\pr}(\k_ir_i)u_l^{\pr}(\k_{i+1}r_{i})
-v_l(\k_ir_i)u_l^\pr(\k_{i+1}r_{i})-
\k_{i+1}r_iv_l(\k_ir_i)u_l^{\pr\pr}(\k_{i+1}r_{i}).
\ea$$
Then one solves problem (\ref{equat}) iteratively:
\be\label{iter}
{\bf\k}^{j+1}={\bf\k}^{j}-
\gamma[{\Phi^\prime}({\bf\k}^{j})]^{-1}\Phi({\bf\k}^{j}),\q
j=0,1,\dots,\q {\bf\k}^{j}:=({\bf\k}_1^j,....,{\bf\k}_N^j),
\ee
where the step $\gamma$ is chosen small enough.
One stops the iterative process when $||\Phi({\bf\k}^{j})||$ becomes smaller than
some small threshold $\epsilon$.

Instead of inverting matrix $\Phi^\prime$ in (\ref{iter}) one denotes 
${\bf h}^j:=[{\Phi^\prime}({\bf\k}^{j})]^{-1}\Phi({\bf\k}^{j})$ and uses the 
Gauss process with the choice of the maximal element by lines and columns 
of the corresponding linear system:
\be
[\Phi^\prime(\k^j)]_{l,m}{\bf h}^j_m=\Phi_l(\k^j),\q
\k^{j+1}=\k^j-\gamma{\bf h}^j.       
\ee
If at a step $l_0$ of the Gauss process, the maximal element of the remaining
elements of the matrix becomes smaller than some threshold $\epsilon_1$ one 
sets all the remaining quantities ${\bf h}^u_l$, $l>l_0$ equal to zero.

\section{Numerical results}          
\vspace*{-0.5pt}
\noindent
\setcounter{equation}{0}
\renewcommand{\theequation}{3.\arabic{equation}}

As was pointed out in the
Introduction, the principal difficulty of the numerical realization
of the inversion schemes, developed for the fixed energy inverse
scattering problem, is
the fast decay of the phase shifts $\d_l$. Therefore one cannot
discriminate between the noise and the data starting from relatively
small $l$.
 So one can use for the numerical reconstruction of a potential
very few phase shifts $\d_0,\dots,\d_{l_0}$,
thus neglecting the rest, which we call the "tail" $\d_l$,
$l>l_0$. 
The goal of our numerical experiments is to show numerically that 
in general the "tail" of the data can not be neglected. 

In the numerical experiments we start 
with some piecewise-constant potential (\ref{pot}) and take
$N=10$. For example let us start with the following potential:
\be
q_i=(1+\cos(i/2))e^{-i},\q i=1,2,\dots,10.
\ee
This potential is denoted in Table 1 by $q_{orig}$. Then, using formulas
(\ref{xk}), (\ref{dkl}) we calculate 
the phase shifts $\d_l$, $l=0,1,\dots,9$,
corresponding to some fixed value 
of $k$ ($k$=2). Then taking the first five phase
shifts $\d_0,\dots,\d_4$ unchanged, we perturb 
the phase shifts $\d_5,\dots,\d_9$ a little, by a random quantity of  
of order of magnitude $\sim 10^{-5}$. This corresponds to the order of
$\d_5$. By the numerical method described in the previous section 
(with $\epsilon_1=10^{-10}$ and $\gamma=1$) we
reconstruct the piecewise-constant potential (denoted in Table 1 as
$q_{rec}$), which generates
the perturbed phase shifts $\tilde{\d_l}$ with the discrepancy 
$\epsilon=10^{-14}$. The corresponding numerical results are 
presented in Table 1 below. 
One can see that
the first five phase shifts $\d_l$ and $\tilde{\d_l}$
of $q_{orig}$ and $q_{rec}$ correspondingly coincide, the remaining five
phase shifts differ by the quatity
of order $10^{-5}$, but 
the potentials $q_{orig}$ and $q_{rec}$ are quite different.        

\vspace{0.5cm}
   
\centerline{Table 1.}

\vspace{0.2cm}

\begin{tabular}{|c|c|c|c|c|}
\hline
&&&&\\
$r_i$ & $q_{orig}$ & $\d_l$ & $\tilde{\d_l}$ & $q_{rec}$\\
&&&&\\
\hline
&&&&\\
$0.5$ & $0.6907240$ & $-9.941752\cdot 10^{-2}$ & 
$-9.941752\cdot 10^{-2}$ & $-2.415259$ 
\\
$1$ & $0.2084572$  & $-3.779873\cdot 10^{-2}$ 
  & $-3.779873\cdot 10^{-2}$  &$1.558406$
\\
$1.5$ & $5.330886\cdot 10^{-2}$  &$-1.179639\cdot 10^{-2}$ 
  &$-1.179639\cdot 10^{-2}$&$-0.589802$
\\
$2$ & $1.069364\cdot 10^{-2}$ & $-3.014222\cdot10^{-3}$ 
 & $-3.014222\cdot10^{-3}$  & $0.355841$
\\
$2.5$ & $1.339883\cdot 10^{-3}$ &$ -6.494566\cdot10^{-4}$ 
& $-6.494566\cdot10^{-4}$  &$-0.171777$  
\\
$3$ & $2.480612\cdot 10^{-5}$ & $-1.691416\cdot10^{-4}$ & 
$-1.719357\cdot10^{-4}$  & $8.157301\cdot10^{-2}$
\\
$3.5$ & $5.794400\cdot 10^{-5}$ & $-1.026825\cdot10^{-4}$  & 
$-9.611266\cdot10^{-5}$ & $-3.368591\cdot10^{-2}$
\\
$4$ & $1.161896\cdot 10^{-4}$ & $-7.547580\cdot10^{-5}$  & 
$-6.558222\cdot10^{-5}$ & $1.191644\cdot10^{-2}$
\\ 
$4.5$ & $9.739553\cdot 10^{-5}$ & $-4.157540\cdot10^{-5}$  & 
$-3.745421\cdot10^{-5}$ & $-3.106548\cdot10^{-3}$
\\ 
$5$ & $5.827817\cdot 10^{-5}$ & $-1.675351\cdot10^{-5}$   
& $-2.219373\cdot10^{-5}$ & $5.785574\cdot10^{-4}$
\\ 
&&&&\\
\hline
\end{tabular}

\vspace{0.5cm}


\end{document}